\documentclass[aps,prb,twocolumn,superscriptaddress, showpacs, byrevtex,floatfix]{revtex4}
\usepackage{graphicx,amsmath}

\begin{document}

\title{Transient terahertz spectroscopy of excitons and unbound carriers\\
in quasi two-dimensional electron-hole gases}

\author{R. A. Kaindl}
\affiliation{Materials Sciences Division, E. O. Lawrence Berkeley National Laboratory and Department of Physics, University of California--Berkeley, Berkeley, CA 94720, USA}

\author{D. H\"{a}gele}
\affiliation{Materials Sciences Division, E. O. Lawrence Berkeley National Laboratory and Department of Physics, University of California--Berkeley, Berkeley, CA 94720, USA}
\affiliation{Arbeitsgruppe Spektroskopie der kondensierten Materie, Ruhr-Universit\"{a}t Bochum, \\D-44780 Bochum, Germany}

\author{M. A. Carnahan}
\affiliation{Materials Sciences Division, E. O. Lawrence Berkeley National Laboratory and Department of Physics, University of California--Berkeley, Berkeley, CA 94720, USA}
\author{D. S. Chemla}
\affiliation{Materials Sciences Division, E. O. Lawrence Berkeley National Laboratory and Department of Physics, University of California--Berkeley, Berkeley, CA 94720, USA}

\begin{abstract}
\noindent We report a comprehensive experimental study and detailed model analysis of the terahertz dielectric response and density kinetics of excitons and unbound electron-hole pairs in GaAs quantum wells. A compact expression is given, in absolute units, for the complex-valued terahertz dielectric function of {\it intra-excitonic} transitions between the 1$s$ and higher-energy exciton and continuum levels. It closely describes the terahertz spectra of resonantly generated excitons. Exciton ionization and formation are further explored, where the terahertz response exhibits both intra-excitonic and Drude features. Utilizing a two-component dielectric function, we derive the underlying exciton and unbound pair densities. In the ionized state, excellent agreement is found with the Saha thermodynamic equilibrium, which provides experimental verification of the two-component analysis and density scaling. During exciton formation, in turn, the pair kinetics is quantitatively described by a Saha equilibrium that follows the carrier cooling dynamics. The terahertz-derived kinetics is, moreover, consistent with time-resolved luminescence measured for comparison. Our study establishes a basis for tracking pair densities via transient terahertz spectroscopy of photoexcited quasi-two-dimensional electron-hole gases.
\end{abstract}

\maketitle

\onecolumngrid \vspace{5 mm} \twocolumngrid

\section{Introduction}\label{Sec_Intro}

\noindent The terahertz (THz) frequency electromagnetic response provides important insight into low-energy excitations and many-body correlations in condensed matter.\cite{Dre02} In semiconductors, Coulomb interactions lead to the formation of excitons from unbound electron-hole ($e$-$h$) pairs. Microscopic interactions on ultrashort time scales determine the dynamics of energy relaxation, dephasing, diffusion, or species inter-conversion of excitons and unbound pairs. Besides opto-electronic applications, understanding the low-energy structure and dynamics of these quasi-particles is fundamentally important, e.g., for the exploration of low-temperature collective phenomena.\cite{Snok02,Buto02,Kasp06,Lai07}

Excitons were extensively studied via optical absorption, photoluminescence (PL), or nonlinear experiments at the semiconductor band gap.\cite{Sha99} Despite their success, these techniques rely on $e$-$h$ pair creation and annihilation which entails important limitations. In particular, momentum conservation often restricts the sensitivity of PL to a subset of excitons around center-of-mass momentum $K \simeq 0$. The intensity then depends not only on the density of the exciton gas but also on its detailed distribution function. Higher-$K$ excitons can be detected in materials with strong electron-phonon coupling, giving insight, e.g., into exciton thermalization and formation.\cite{Uml98,Haeg99,OHar00} However, the determination of {\it absolute} densities is exceedingly difficult lacking precise knowledge of the collection efficiency and inter-band dipole moment.

In contrast, {\it intra-excitonic} THz transitions between the low-energy internal levels of excitons represent a fundamentally different tool. They measure the coupling between the 1$s$ exciton ground state and higher relative-momentum states, detecting excitons largely independent of $K$. Intra-excitonic probes can determine {\it absolute} exciton densities, since they measure -- analogous to atomic absorption spectroscopy -- existing excitons with predictable cross section. These THz transitions are independent of the interband dipole moment, which renders them suitable probes in the search for exciton condensation. Importantly, THz fields are also equally sensitive to unbound $e$-$h$~pairs (free carriers).

While initial studies were scarce, intra-excitonic spectroscopy recently emerged as a powerful tool to investigate the low-energy resonances and dynamics of excitons, \cite{Hak58,Tim76,Nik84,Gro94,Cer96,John01,Kir01,Kai03,Gono04,Kub05,Karp05,Jorg05,Gal05,Hub05,Hub06a,Taya06,Hend07,Hugh08,Ideg08}
fueled in part by rapid advances in ultrafast THz technology.\cite{Dex07} Optical-pump THz-probe experiments are of particular interest, as they yield both real and imaginary parts of the transient response functions. Notably, we reported distinct THz signatures of conducting and insulating phases during exciton formation and ionization in GaAs quantum wells  (QWs).\cite{Kai03} Such complex-valued THz spectra place strict boundaries on theoretical models, providing exceptional potential for further analysis to directly determine exciton and free carrier densities.

Here, we report a detailed model analysis and experimental study of the transient THz spectra and density kinetics of quasi-two-dimensional (2D) excitons and unbound $e$-$h$ pairs in GaAs QWs. An accurate evaluation procedure for the optical-pump THz-probe signals is described for the multi-layer QW geometry. A key aspect of this work is the derivation of the intra-excitonic dielectric function scaled in absolute units, as a compact expression to determine pair densities from measured THz spectra. It closely describes both shape and amplitude of the THz response during population decay of resonantly-generated excitons. In contrast, the THz line shapes during exciton ionization and formation exhibit both intra-excitonic and Drude-like features. A two-component dielectric function is applied to infer the underlying exciton and free-carrier fractions. After ionization, this reveals a mixture that agrees closely with the thermodynamic equilibrium (Saha) predictions -- thus experimentally verifying the two-component analysis. Its application to exciton formation reveals a surprisingly simple yet quantitative description of the dynamics, when cooling of the $e$-$h$ gas is taken into account. This THz-derived scenario is found to consistently describe the luminescence dynamics measured for comparison. Hence, this paper establishes a basis for gauging exciton and free-carrier densities in the THz dielectric response of quasi-2D $e$-$h$ gases, and demonstrates its application to tracing the time evolution of multi-component phases during exciton ionization and formation.

In the following, Sec.~\ref{Sec_Methods} explains the experimental methods, while the THz response of resonantly-generated excitons and unbound pairs is discussed in Sec.~\ref{Sec_ResHH}. Models of the intra-excitonic and Drude response are presented that enable determination of absolute pair densities. Section \ref{Sec_Ioniz} discusses the THz response and kinetics during exciton ionization at elevated lattice temperatures. In Sec.~\ref{Sec_Form}, exciton formation after non-resonant excitation is analyzed and compared to PL kinetics. Appendixes \ref{Sec_EvalAppendix} and \ref{Sec_ExcitonTheoryAppendix} derive the complex transmission function of the quantum-well structure, and the intra-excitonic dielectric function, respectively.

\section{Experimental Technique}\label{Sec_Methods}

\begin{figure}[t!]      
\includegraphics[width=7.5cm]{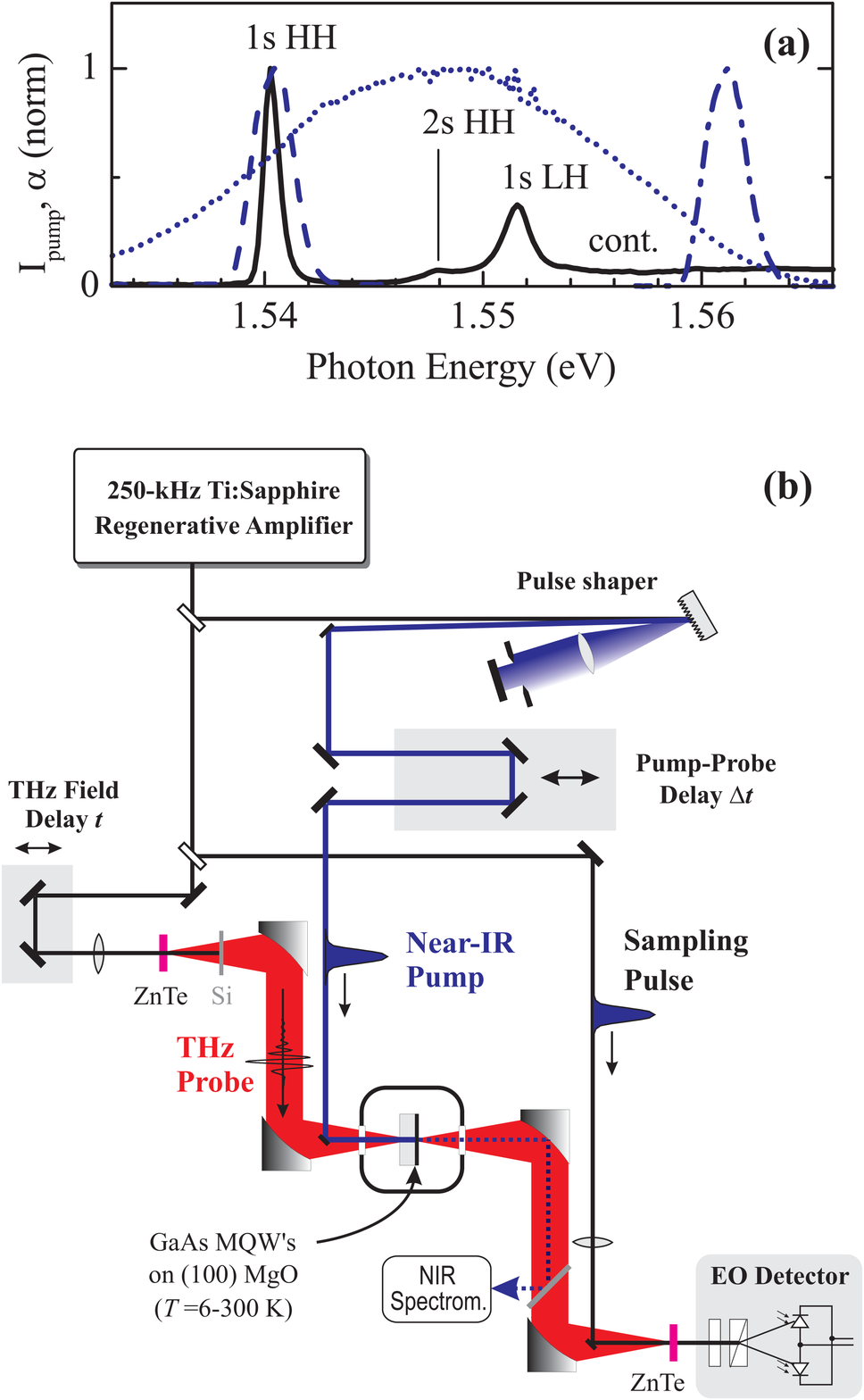}%
\caption{(color online). (a) Near-IR absorption of the GaAs multi-quantum-well sample at $T = 6$~K (solid line). A typical laser spectrum is shown unshaped (dotted line), and shaped for resonant 1$s$-HH excitation (dashed line) or continuum excitation (dashed-dotted line). (b) Experimental setup for optical-pump THz-probe spectroscopy. The THz section is purged with dry nitrogen to avoid far-infrared absorption in air.
\label{Fig_QWAbs_Setup}}
\end{figure}

\begin{figure*}[t!]          
\includegraphics[width=17.8cm]{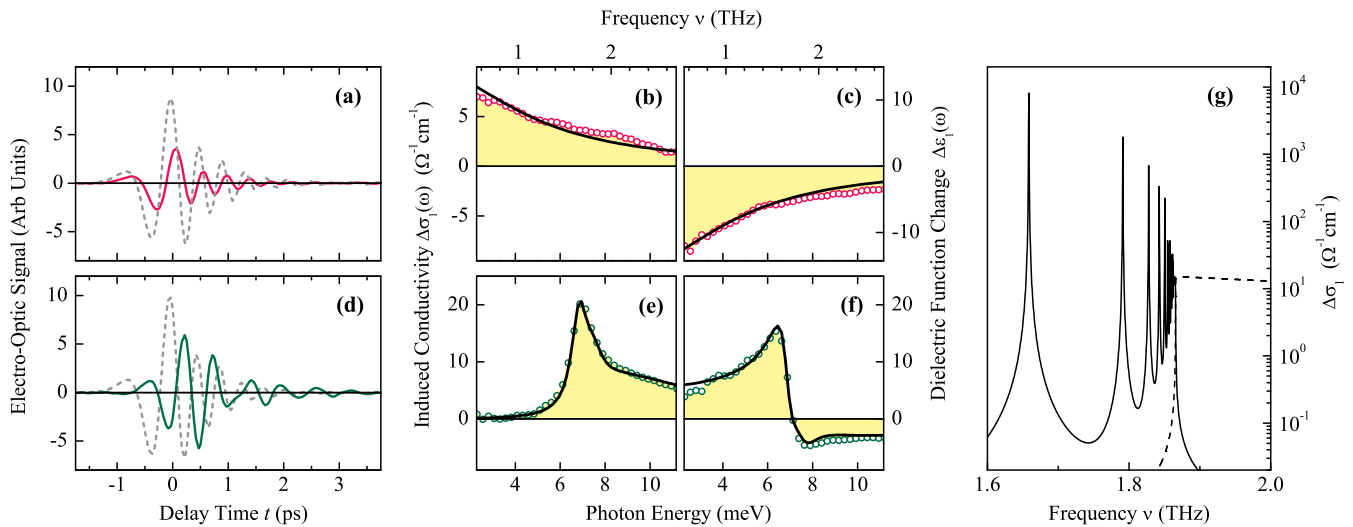}
\caption{(color online). [(a)-(c)] THz response at $T_{\rm L} = 300$~K, at delay $\Delta t = 10$~ps after {\it non-resonant} excitation 100 meV above the band gap with excitation fluence $F = 0.6~\mu$J/cm$^2$. The THz transients in (a) are the reference (dashed line) and pump-induced change (solid line, multiplied $\times 50$). Panels (b) and (c) show corresponding spectra $\Delta\sigma_1$ and $\Delta\epsilon_1$ from experiment (dots), and a Drude model (solid line) with $n_{\rm eh} = 2 \times 10^{10}$~cm$^{-2}$ and $\Gamma_{\rm D} = 4.8$~meV. [(d)-(f)] THz response for {\it resonant} $1s$-HH excitation at $T_{\rm L} = 6$~K, with $F = 0.14~\mu$J/cm$^2$ and $\Delta t$~=~10~ps. Solid lines in panels (e) and (f): intra-excitonic model, $n_{\rm X} = 2.7\times10^{10}$~cm$^{-2}$. (g) Intra-excitonic model as in panel (e), but $\Gamma_{\rm bb} = \Gamma_{\rm bc} = 1~\mu$eV. Solid line: $1s$-n$p$ transitions; dashed line: $1s$-continuum.
\label{Fig_THzDrudeExciton}}
\end{figure*}

\noindent In the experiments, we utilize optical-pump THz-probe spectroscopy to investigate transient changes in the THz conductivity of photoexcited $e$-$h$ gases in GaAs QWs. The high-quality sample studied here consists of a stack of ten 14-nm~wide, undoped GaAs wells separated by 10-nm~wide Al$_{0.3}$Ga$_{0.7}$As barriers, grown via molecular beam epitaxy on a GaAs substrate.\cite{Lov02} The structure is embedded in 500~nm-thick Al$_{0.3}$Ga$_{0.7}$As spacer layers. To avoid THz absorption of photoexcited carriers in the substrate, it was removed by selective etching after attaching the QW side to a 0.5-mm thick $\left<100\right>$ MgO substrate.\cite{Note002} The corresponding low-temperature near-IR absorption spectrum is shown by the solid line in Fig.~\ref{Fig_QWAbs_Setup}(a). The $1s$ heavy-hole (HH) exciton line at 1.540~eV dominates the spectrum, with a line width of 0.8~meV [full width at half maximum (FWHM)]. The $2s$~HH and $1s$~light-hole (LH) absorption lines, and the inter-band continuum follow at higher photon energies.

The experimental setup is shown in Fig.~\ref{Fig_QWAbs_Setup}(b). We utilize amplified pulses at high repetition-rate to allow for both sensitive THz detection and sufficiently intense pulses to excite the intrinsically large THz probe area. At the outset, a 250-kHz Ti:sapphire amplifier system (Coherent RegA) delivers 150-fs near-IR pulses at 800~nm wavelength. A fraction of the output is used to generate THz probe pulses via optical rectification and to detect them via electro-optic sampling, each in a 500-$\mu$m thick $\left<110\right>$ ZnTe crystal.\cite{Kai01a} The THz pulses span the 2--12~meV ($\approx$~0.5--3~THz) spectral range. The THz beam is recollimated and focused with off-axis parabolic mirrors onto the sample mounted in a cold-finger cryostat. For optimal time resolution and to avoid spectral pump-probe artifacts, the THz field is scanned via the generation pulse delay and the pulses are incident from the substrate side.\cite{Bea00} The THz focus size is frequency-dependent, with a FWHM diameter of around 1~mm at 1.5 THz.\cite{Note001} We employ a 2-mm diameter aperture to limit the THz probe to the photoexcited region.

A second part of the near-IR amplifier output is used for photoexcitation. The full laser spectrum is shown as the dotted line in Fig.~\ref{Fig_QWAbs_Setup}(a). For selective excitation of excitons or unbound pairs, the pulses must be spectrally shaped. This is achieved in a reflective, zero-dispersion stretcher consisting of a 1200 l/mm grating, 200-mm focal length lens, and an adjustable slit. Pump spectra are narrowed to $\approx 1$--$2$~meV width, with typical curves shown as the dashed and dashed-dotted curves in Fig.~\ref{Fig_QWAbs_Setup}(a). This yields an overall 1-2 ps time resolution. On-line spectral characterization of the pump light transmitted through the sample allows for precise determination of the exciton line position, for optimal spectral overlap and resonant excitation at varying sample temperatures.

The THz studies are complemented by time-resolved PL on the identical sample. Such data were taken with the sample mounted in a vapor-flow cryostat and photoexcited by a 76-MHz Ti:sapphire oscillator, spectrally narrowed using a 1-meV wide interference filter. Linearly polarized excitation was used to minimize spin effects. The luminescence was temporally resolved with a Hamamatsu streak camera. These measurements necessitated a smaller pump spot diameter ($140 \mu$m), which however remains well beyond estimated carrier diffusion lengths within the 1-ns time window.

In the THz experiments, the dynamics of the dielectric response is determined as follows. After photoexcitation, the THz dielectric function of the QW layers in equilibrium, denoted by $\epsilon(\omega)$, transiently changes to the modified value $\epsilon(\omega) + \Delta\epsilon(\omega)$. Thus, the induced change $\Delta\epsilon(\omega)$ must be determined for each fixed time delay $\Delta t$ between the arrival of pump and probe pulses on the sample. For this, we measure the THz reference probe field $E(t)$ (transmitted through the sample in equilibrium) and its pump-induced change $\Delta E(t)$. Typical time-domain THz signals are shown in Figs.~\ref{Fig_THzDrudeExciton}(a) and \ref{Fig_THzDrudeExciton}(d). Fourier transformation provides the corresponding frequency-domain fields $E(\omega)$ and $\Delta E(\omega)$. The change in the dielectric response is then obtained as a function of these fields,

\begin{equation}
\Delta\epsilon(\omega) = f\!\!\left(\frac{\Delta E(\omega)}{E(\omega)+\Delta E(\omega)}\right)
\end{equation}

\noindent which depends on the sample geometry. An analytical expression $f$ that takes into account the multilayer structure of our QW sample is derived in Appendix \ref{Sec_EvalAppendix} [Eq.~(\ref{Eq_Dn_multi})].

The current response $J(\omega) = \sigma(\omega)E(\omega)$ of the many-particle system to the incident transverse electromagnetic field is given by the optical conductivity $\sigma(\omega)= \sigma_1(\omega) + i\sigma_2(\omega)$. It is connected to the dielectric function via $\sigma(\omega) = i\omega\epsilon_0[1-\epsilon(\omega)]$. In the following, we will express the transient THz response as
\begin{equation}
\Delta\epsilon(\omega) = \Delta\epsilon_1 (\omega) + \frac{i}{\epsilon_0 \omega} \Delta\sigma_1 (\omega).
\end{equation}
\noindent  Here, the induced conductivity $\Delta\sigma_1(\omega)$ is a measure of the absorbed power density and allows for analysis of oscillator strengths. The dielectric function change $\Delta\epsilon_1(\omega)$, in turn, provides a measure of the inductive, out-of-phase response. As evident below, the availability of both $\Delta\sigma_1$ and $\Delta\epsilon_1$ is a key to distinguishing different contributions to the multi-component THz spectra.

\section{Terahertz Response of e-h Pairs}\label{Sec_ResHH}

\noindent
We first discuss experiments that probe transient changes in the THz dielectric response of unbound $e$-$h$ pairs after {\it non-resonant} photoexcitation into the band-to-band continuum. Data are shown in Figs.~\ref{Fig_THzDrudeExciton}(a)-\ref{Fig_THzDrudeExciton}(c) for lattice temperature $T_{\rm L} = 300$~K, which ensures rapid ionization of the $e$-$h$ pairs. As evident in the time-domain THz traces in Fig.~\ref{Fig_THzDrudeExciton}(a), the pump-induced field change (solid line) resembles the reference (dashed line) with a phase shift, pointing to a spectrally broadband response. This is confirmed by $\Delta\sigma_1(\omega)$ and $\Delta\epsilon_1(\omega)$ shown as dots in Figs.~\ref{Fig_THzDrudeExciton}(b) and \ref{Fig_THzDrudeExciton}(c). The large low-frequency conductivity $\Delta\sigma_1$ underscores the conducting nature of the unbound pairs, while the dispersive $\Delta\epsilon_1 < 0$ is characteristic of a zero-frequency, Drude-like oscillator. Indeed, the response is well described by the Drude model (solid lines)
\begin{equation}
\Delta\epsilon(\omega) = n_{\rm eh}\Delta\epsilon_{\rm D}(\omega) = n_{\rm eh}\cdot\frac{-e^2}{d_{\rm W}\epsilon_0 \mu \,(\omega^2+i\omega\Gamma_{\rm D})}~,
\end{equation}
\noindent where $n_{\rm eh}$ is the $e$-$h$ pair sheet density per well, $d_{\rm W}$ is the QW width, $\Gamma_{\rm D}$ is the Drude scattering rate, and $\mu \equiv m_{\rm e} m_{\rm h}/(m_{\rm e}+m_{\rm h})$ is the reduced mass ($m_{\rm e}, m_{\rm h}$ are $e$ and $h$ effective masses, respectively). The density $n_{\rm eh}$ from the Drude fit closely agrees with the value ($1.9\times10^{10}$~cm$^{-2}$) estimated from the excitation fluence and sample parameters.

Next, we will discuss the transient THz response after {\it resonant} excitation at the $1s$-HH exciton line, for $T_{\rm L} = 6$~K. The measured THz field change in Fig.~\ref{Fig_THzDrudeExciton}(d) exhibits a complex shape with time-dependent phase shift. Figures~\ref{Fig_THzDrudeExciton}(e) and \ref{Fig_THzDrudeExciton}(f) show the corresponding THz spectra (dots). Here, the conductivity $\Delta\sigma_1$ is characterized by a distinct, asymmetric peak around $\hbar\omega\approx$~7~meV. The dielectric function change $\Delta\epsilon_1$, in turn, shows an oscillatory response around the same photon energy. This represents a new low-energy oscillator, absent in equilibrium, which can be explained by transitions between an exciton's \textit{internal} degrees of freedom. The 7~meV conductivity peak arises from the 1$s$~$\rightarrow$~2$p$ transition between the exciton levels, in concordance with GaAs/AlGaAs QW exciton binding energies\cite{Ger98} and the $1s-2s$ splitting in Fig.~\ref{Fig_QWAbs_Setup}(a). The vanishing low-frequency conductivity $\Delta\sigma_1$ of the intra-excitonic response is a signature of the insulating nature of the charge-neutral excitons. Compared to the Drude response, the opposite sign of $\Delta\epsilon_1$ at low frequencies enables further discrimination between the THz response of excitons and unbound pairs.

For a quantitative description we have performed calculations, scaled in absolute units, of the intra-excitonic contribution to the dielectric function. The model takes into account 2D bound and continuum hydrogenic wave functions, where the Bohr radius
\begin{equation}
a \equiv \frac{4 \pi \hbar^2 \epsilon_0 \epsilon_{\rm s}}{e^2\mu\lambda}
\label{Eq_bohr_radius}
\end{equation}
\noindent and binding energies
\begin{equation}
E_n = -\frac{e^2\lambda}{8 \pi \epsilon_0 \epsilon_{\rm s} a}\cdot(n+1/2)^{-2}
\label{Eq_binding_energy}
\end{equation}
\noindent are scaled by the reduced mass $\mu$ and static dielectric constant $\epsilon_{\rm s}$, and by  a parameter $\lambda$ which scales the Coulomb potential to take into account the finite well width.\cite{Eke87} The complex-valued dielectric THz response from intra-excitonic transitions between $1s$ and higher bound and continuum states is then given by
\begin{widetext}
\begin{eqnarray}\label{Eq_IntraExcitonDielecFunc}
\Delta\epsilon(\omega) & = & n_{\rm X} \Delta\epsilon_{\rm X}(\omega) = n_{\rm X} \frac{2 e^2 a^2}{d_{\rm W} \hbar^2 \epsilon_0} \{
\sum_{n=1}^{\infty} \frac{E_n-E_0}{([\frac{E_n-E_0}{\hbar}]^2-\omega^2)-i\omega\Gamma_{\rm bb}}
~\cdot(1-\frac{1}{2n+2})^5(1-\frac{1}{n+1})^{2n}~n^{-3} + \\
\nonumber & & \hspace{26 mm} \int_0^{\infty}
\frac{E_0 + E(k)}{([\frac{E_0 + E(k)}{\hbar}]^2-\omega^2)-i\omega\Gamma_{\rm bc}}
~\cdot\frac{a^2 k}{1+e^{-\frac{2\pi}{ak}}} ~\frac{(\frac{2i+ak}{2i-ak})^{-\frac{2i}{ak}}}{(1+[ak/2]^2)^4}~dk \}
\end{eqnarray}
\end{widetext}
\noindent where $n_{\rm X}$ is the $1s$ exciton sheet density, $\Gamma_{\rm bb}$ and $\Gamma_{\rm bc}$ are level broadenings for bound-bound and bound-continuum transitions, and $E(k) = \hbar^2k^2/2\mu$ is the kinetic energy of the continuum states. Details of the derivation are given in Appendix~\ref{Sec_ExcitonTheoryAppendix}. To illustrate the underlying transitions and oscillator strengths, Fig.~\ref{Fig_THzDrudeExciton}(g) shows a calculated intra-excitonic line shape for (unrealistically low) 1~$\mu$eV broadening. It consists of the $1s$-$2p$ peak and transitions into higher bound $n$p levels (solid line), and transitions into the continuum of unbound $e$-$h$ pairs (dashed). As for three-dimensional intra-atomic transitions, bound-bound and bound-continuum transitions match up smoothly at the continuum edge.

We now compare this model with the experimental data. The calculated intra-excitonic response is shown as solid lines in Fig.~\ref{Fig_THzDrudeExciton}(e) and \ref{Fig_THzDrudeExciton}(f), which reproduces the shape of the experimental data extremely well. The sharp lines are now absent due to realistic broadening, and the response is dominated by the $1s$-$2p$ peak and higher-energy shoulder. Most of the model parameters are severely restricted: for the GaAs QWs $\epsilon_{\rm s} = 13.2$ and $\mu = 0.054 m_0$ ($m_0$: free electron mass), using $m_{\rm e} = 0.0665 m_0$ and $m_{\rm h} = 0.28 m_0$ averaged in $k$-space over $\approx \,1/a$ to account for the exciton wave function.\cite{Sia00} Moreover, $\lambda = 0.678$ must be chosen to reproduce the observed $1s$-$2p$ level spacing. This leaves, as the only free parameters, the $1s$ exciton density $n_{\rm X}$ and level broadenings $\Gamma$. The best agreement in shape in Figs.~\ref{Fig_THzDrudeExciton}(e) and \ref{Fig_THzDrudeExciton}(f) is obtained with $\Gamma_{\rm bb} = 0.8$~meV for bound-bound transitions and $\Gamma_{\rm bc} = 2.2$~meV for bound-continuum transitions. The larger $\Gamma_{\rm bc}$ reflects increased scattering of continuum final states, as corroborated by a comparable Drude width of the non-resonantly excited, $T = 6$~K response discussed further below. For the experimental data in this paper, the consistently best fit was obtained by keeping $\Gamma_{\rm bc}$ fixed while varying the intra-excitonic broadening $\Gamma_{\rm bb}$. The 1$s$ exciton density from the model fit ($n_{\rm X} = 2.7\times10^{10}$~cm$^{-2}$) in Figs.~\ref{Fig_THzDrudeExciton}(e) and \ref{Fig_THzDrudeExciton}(f) compares well with the density $2.1\times10^{10}$~cm$^{-2}$ estimated from the pump flux ($0.14~\mu$J/cm$^2$) in the experiment after accounting for sample absorption, spectral overlap, and reflection losses of the cryostat windows and sample. Work at densities well below the present $10^{10}$~cm$^{-2}$ range is desirable, but will necessitate further improvements in the sensitivity of the THz measurement technique.

\begin{figure}          
\includegraphics[width=7cm]{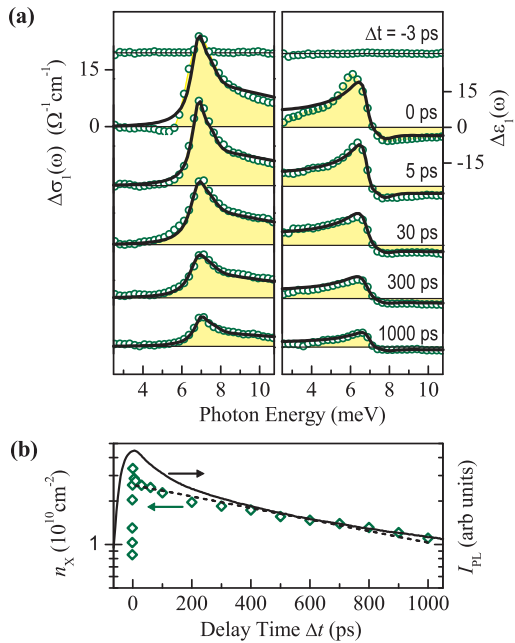}
\caption{(color online).
(a) Transient THz response (dots) for different delays $\Delta t$, after resonant $1s$-HH excitation as in Figs.~\ref{Fig_THzDrudeExciton}(e) and \ref{Fig_THzDrudeExciton}(f). Lines: intra-excitonic dielectric model, with $\Gamma_{\rm bb}~= 0.9$~meV (0~ps), 0.8~meV (5~ps), 1.03~meV (30~ps), 1.1~meV (300~ps), and 0.9~meV (1000~ps). Curves shifted vertically are scaled identically. (b) Exciton density $n_{\rm X}$ (diamonds) from the model in panel (a). Dashed line: exponential decay with $\tau = 1087$~ps. Solid line: PL intensity after resonant excitation, for $n~\simeq~2\times10^{10}$~cm$^{-2}$ and $T_{\rm L}~=~10$~K. \label{Fig_Recombination}}
\end{figure}

With respect to absolute density scaling, it should be emphasized that both the Drude and intra-excitonic models above fulfill the ``partial oscillator strength sum rule''
\begin{equation}
     \int_0^\infty \sigma_1(\omega) d\omega = \frac{\pi}{2} \frac{ne^2}{d_{\rm W} \mu} \;,
\end{equation}
\noindent as expected in a parabolic band approximation. The total photoexcited sheet density $n$ of bound and unbound pairs can thus be directly obtained from the integral of $\Delta \sigma_1$ (to below the onset of inter-band transitions), which corresponds to the induced intra-band spectral weight. This underscores the capability of THz spectroscopy to determine, unlike luminescence, absolute densities of excitons and unbound $e$-$h$ pairs both at $K = 0$ and outside the optically-accessible momentum range.

Transient THz spectra at several different pump-probe delays are shown in Fig.~\ref{Fig_Recombination}(a). Initially, a coherent $1s$ exciton polarization is created by the near-IR pump pulse which dephases within a few picoseconds into an incoherent exciton population.\cite{Lov02} In the coherent regime directly after excitation, the THz response deviates noticeably from the model line shape, as evident at $\Delta t = 0$~ps in Fig.~\ref{Fig_Recombination}(a). At later times ($\Delta t \gtrsim 5$~ps) the intra-excitonic dielectric function well describes the transient THz response. The conductivity decays in amplitude but retains its peaked line shape, evidencing directly the decay of excitonic populations. Changes in the broadening $\Gamma_{\rm bb}$ are minor, and may result from a time-varying temperature of the exciton gas due to recombination heating. Densities derived from the model fits are charted in Fig.~\ref{Fig_Recombination}(b) (diamonds). An exciton recombination time $\tau \simeq 1$~ns is obtained from the single-exponential fit (dashed line). The radiative decay of free $K = 0$ excitons in QWs is predicted to be as short as 10~ps in the idealized case without any dephasing. PL experiments on a highest-quality single-QW indeed exhibited decays as short as 40~ps, for narrow well width (4.5~nm), low density ($3\times10^9$~cm$^{-2}$) and very low temperature (1.7~K).\cite{Dev91} However, for comparison with the THz-derived dynamics we must consider the case of wide QWs and densities exceeding $10^{10}$~cm$^{-2}$, where PL decay times are much longer and consistent with our result.\cite{Fel87,Mar93}

We also compare the THz results to our time-resolved PL measurements after resonant 1$s$-HH excitation, indicated by the solid line in Fig. \ref{Fig_Recombination}(b). The PL exhibits an initially faster decay, which can arise from unsuppressed spin relaxation effects, coherent emission, or pump scattering, effects that typically complicate luminescence transients. At later times the PL agrees well with the THz dynamics. Unlike the THz experiment, however, absolute densities are intrinsically difficult to obtain from the PL intensity, despite that in this resonantly-excited case it reflects recombination of mainly cold excitons in $K\approx0$ luminescent states.

\section{Exciton Ionization}\label{Sec_Ioniz}

\begin{figure*}         
\includegraphics[width=17.5cm]{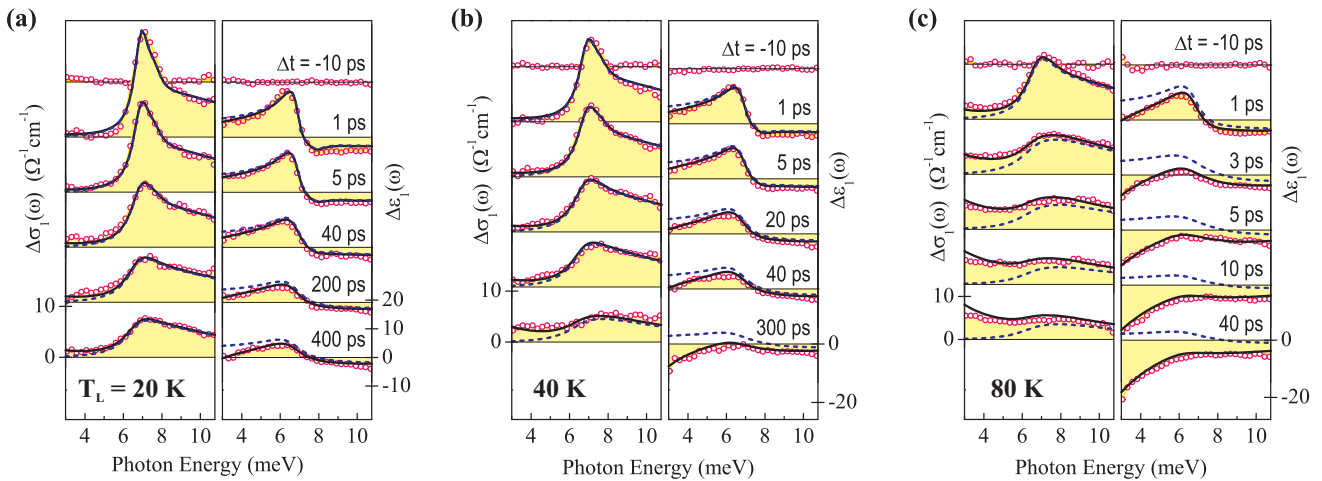}
\caption{(color online). Transient THz spectra at lattice temperatures (a) $T_{\rm L} =$ 20~K, (b) 40~K, and (c) 80~K after resonant $1s$-HH excitation with fluence of $0.14~\mu$J/cm$^2$. Dots: induced conductivity and dielectric function change at indicated pump-probe delays $\Delta t$. Curves are shifted vertically, but scaled equally. Solid lines: two-component model; dashed lines: intra-excitonic model only, with time-varying density and broadening.\cite{Note003}
\label{Fig_TempDep1}}
\end{figure*}

\begin{figure}[t]           
\includegraphics[width=7.6cm]{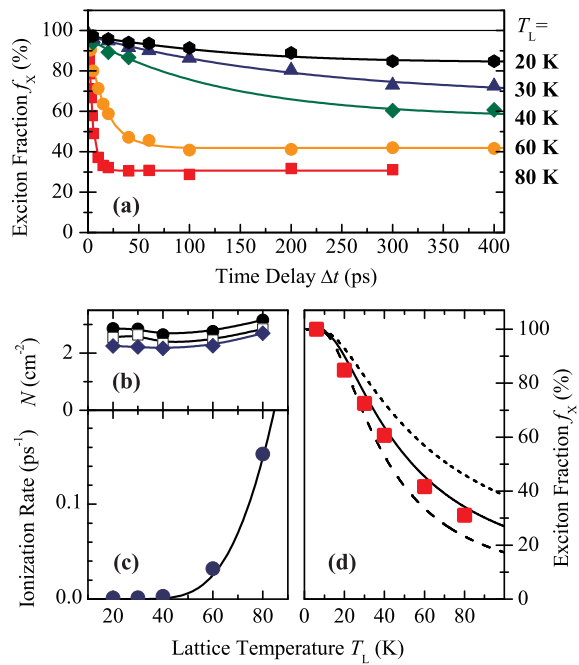}
\caption{(color online). (a) Ionization kinetics after resonant excitation at indicated temperatures. Symbols: exciton fraction $f_{\rm X}$ from the two-component analysis. Lines: fit with exponential decay plus offset. (b) Total pair density $N$ for $\Delta t$ = 5 ps (dots), 40 ps (squares), and 300~ps (diamonds). (c) Exciton ionization rate (dots) from the initial decay of $f_{\rm X}$ compared to LO-phonon scattering (line) with $\Gamma_{\rm LO} = 30$~THz. (d) Exciton fraction $f_{\rm X}$ at long times (squares), compared to the Saha model for $N=2\times10^{10}$~cm$^{-2}$ (solid line), $1\times10^{10}$~cm$^{-2}$ (dashed line), and $4\times10^{10}$~cm$^{-2}$ (short-dashed line).\label{Fig_Ionization Analysis}}
\end{figure}

\noindent At increased lattice temperature, the temporal dynamics and shape of the transient THz spectra undergo extensive changes. Figure~\ref{Fig_TempDep1} shows the response after resonant HH excitation, for three representative temperatures. At 20~K, $\Delta\sigma_1(\omega)$ is well described by a sharp exciton line shape directly after excitation, but broadens noticeably with increasing time delay [Fig.~\ref{Fig_TempDep1}(a)]. Simultaneously, $\Delta\epsilon_1(\omega)$ flattens out. As shown in Figs.~\ref{Fig_TempDep1}(b) and \ref{Fig_TempDep1}(c), these changes occur faster and become even more enhanced as the lattice temperature is further elevated to 40 and 80~K. Two important hallmarks of a Drude response appear: at the {\it low-frequency} end of the spectrum significant conductivity $\Delta\sigma_1$ builds up, and the induced dielectric function $\Delta\epsilon_1$ increasingly deviates towards negative values. The dynamics in Fig.~\ref{Fig_TempDep1} thus evidences the generation of unbound $e$-$h$ pairs, which indicates thermal ionization of the resonantly excited HH excitons.

\subsection{Two-component analysis}

To obtain a quantitative picture of the ionization process, we need to describe the complex spectra at all delay times. The intra-excitonic model function alone (dashed lines in Fig.~\ref{Fig_TempDep1}) is clearly insufficient. Indeed, the intra-excitonic $\Delta\epsilon_1$ always remains positive below the 1$s$-2$p$ oscillator frequency ($\nu \lesssim 1.7$~THz), regardless of the amount of broadening. This underscores the importance of measuring both real and imaginary parts of the THz response. In order to take into account the simultaneous existence of excitons and unbound $e$-$h$ pairs, we implement a two-component dielectric function
\begin{equation}
\Delta\epsilon(\omega) = n_{\rm X} \Delta\epsilon_{\rm X}(\omega)
+ n_{\rm eh} \Delta\epsilon_{\rm D}(\omega)~,
\end{equation}
\noindent where $\Delta\epsilon_{\rm X}$ is the intra-excitonic dielectric function and $\Delta\epsilon_{\rm D}$ is the Drude response described above. This model was fitted to each THz spectrum by varying the densities $n_{\rm eh}$, $n_{\rm X}$ and the broadening parameters. Resulting model functions are shown as solid lines in Fig.~\ref{Fig_TempDep1}. They describe the experimental data well, while neither an exciton nor a Drude model alone can account for the response at long delay times and elevated temperatures. The fit parameters are strongly constrained due to important spectral differences between the response of excitons and unbound $e$-$h$ pairs, and by the need to explain both $\Delta\sigma_1(\omega)$ and $\Delta\epsilon_1(\omega)$ simultaneously and over a broad spectral range. These THz spectra are reproduced with a fixed binding energy in the model; we verified that, using the equations in Ref.~\onlinecite{Sno08}, that renormalization due to free-carrier screening should indeed remain $\lesssim\!\!10$\% for our conditions. However, a distinct time-dependent broadening is observed which reflects changes in homogeneous dephasing of both the 1$s$ and 2$p$ levels due to exciton--free carrier and exciton--phonon interactions, with possibly inhomogeneous contributions due to a momentum-dependent binding energy.\cite{Sia00}

With knowledge of $n_{\rm eh}$ and $n_{\rm X}$ in absolute units, we can obtain the exciton fraction
\begin{equation}
f_{\rm X} \equiv \frac{n_{\rm X}}{n_X + n_{\rm eh}},
\end{equation}
as a measure of the admixture of bound $e$-$h$ pairs to the many-particle system. Figure~\ref{Fig_Ionization Analysis}(a) shows the temporal dynamics of $f_{\rm X}$ obtained from two-component fits to the experimental data. At higher lattice temperatures, $f_{\rm X}$ decays with time until it reaches a {\it quasi-equilibrium} value. As evident, with rising lattice temperature the ionization becomes faster while the residual, quasi-equilibrium exciton fraction at long delay times decreases. The total pair density is fairly constant for all temperatures, as shown in Fig.~\ref{Fig_Ionization Analysis}(b), and decays monotonously with time.

Figure~\ref{Fig_Ionization Analysis}(c) shows the temperature dependent ionization rate, as derived from the initial decay rate of the exciton fraction, $\partial f_{\rm X} / \partial t |_{t=0}$ from Fig.~\ref{Fig_Ionization Analysis}(a). We can explain the large temperature dependence above $T_{\rm L}\gtrsim$~50~K by ionization through LO phonon absorption. The scattering rate of this process is given by $\Gamma = \Gamma_{\rm LO} \cdot n(T_{\rm L})$, where $n(T) \equiv (\exp(\Omega_{\rm LO}/k_{\rm B}T)-1)^{-1}$ is the Bose occupation with $\Omega_{\rm LO} = 36.6$~meV for GaAs. To describe the experiment, a corresponding model function with $\Gamma_{\rm LO} = 30$~THz is shown in Fig.~\ref{Fig_Ionization Analysis}(c) (solid line). This value for $\Gamma_{\rm LO}$ is in excellent agreement with the LO phonon scattering rates of excitons derived from the broadening of the near-IR exciton absorption lines.\cite{Gamm95_comment}

\subsection{Thermodynamic quasi-equilibrium}

We can now compare the quasi-equilibrium observed at long delay times to the predictions of the so-called Saha equation. The latter describes the densities of excitons and free carriers after statistical equilibration of their chemical potentials in the Boltzmann limit. For a 2D gas of $e$-$h$ pairs, it reads:\cite{Che84,Colo90,Yoo96}
\begin{equation}
\frac{(N - n_X)^2}{n_{\rm X}} = \frac{k_{\rm B} T}{2 \pi \hbar^2} ~\mu ~e^{-E_0/k_{\rm B} T}, \label{Eq_Saha}
\end{equation}
\noindent where $N \equiv  n_{\rm eh} + n_{\rm X}$ is the total e-h pair density. The reduced mass $\mu = 0.054 m_0$ and binding energy $E_0 = 7.7$~meV are retained from the above lineshape model. For a given total pair density $N$, Eq.~\ref{Eq_Saha} then yields the temperature dependence of $n_{\rm X}$, and hence $f_{\rm X}(T)$, in the thermodynamic equilibrium.

Figure~\ref{Fig_Ionization Analysis}(d) compares the Saha prediction of $f_{\rm X}(T)$ (lines) with the experimentally-derived values (squares) for long delay times. A pair density $N~=~2~\times~10^{10}$~cm$^{-2}$ (solid line) yields surprisingly close agreement with the temperature dependence of the experimentally-determined exciton fraction. This extends previous PL-based studies of the 100--300~K range (Ref.~\onlinecite{Colo90}) to temperatures below 100~K and provides quantitative density information accessible only to THz probes. Importantly, not only the shape $f_{\rm X}(T)$ but also the {\it absolute} density $N$ underlying this Saha model curve agrees well with the values obtained from the THz spectra, as evident from the total pair density at long delay times ($\Delta t$~=~300~ps) shown as diamonds in Fig.~\ref{Fig_Ionization Analysis}(b).

An alternate calculation of the THz response of mixed e-h gases was also reported,\cite{Kira04b} which for THz spectra with overwhelming 90\% intra-excitonic oscillator strength predicts an exciton fraction of only 10\%.\cite{Ste08} Consider this picture applied, e.g., to our THz spectra after ionization at $T=20$~K corresponding to $f_{\rm X} = 85$\% [Figs.~\ref{Fig_TempDep1}(a) and \ref{Fig_Ionization Analysis}(d)]. The model of Ref.~\onlinecite{Ste08} accordingly implies an exciton fraction $<10$\%, more than six times the ionization level possible by thermal excitations at the given temperature, density, and binding energy. Also, since 1$s$-2$p$ THz absorption directly measures exciton populations,\cite{Kir01} the exciton density is $\simeq2\times10^{10}$~cm$^{-2}$ [Fig.~\ref{Fig_TempDep1}(a), 400~ps] using our gauge from Sec.~\ref{Sec_ResHH}. A 10\% exciton fraction would then imply a total pair density of $2\times10^{11}$~cm$^{-2}$, {\it exceeding} the known, absorbed photon density more than seven times. Given these stark discrepancies, we cannot employ the model of Refs.~\onlinecite{Kira04b}~and~\onlinecite{Ste08} for quantitative insight into THz spectra of mixed $e$-$h$ gases.

In contrast, the self-consistent, quantitative agreement shown in Fig.~\ref{Fig_Ionization Analysis} between $(i)$ the Saha-model densities and exciton fraction founded on basic thermodynamic relations, and $(ii)$ the experimentally-derived total pair density and exciton fraction from our analysis during the quasi-equilibrium at long delay times provides a clear and direct validation of our two-component dielectric function analysis of the transient THz spectra.

\section{Exciton Formation}\label{Sec_Form}

\begin{figure}          
\includegraphics[width=7.5cm]{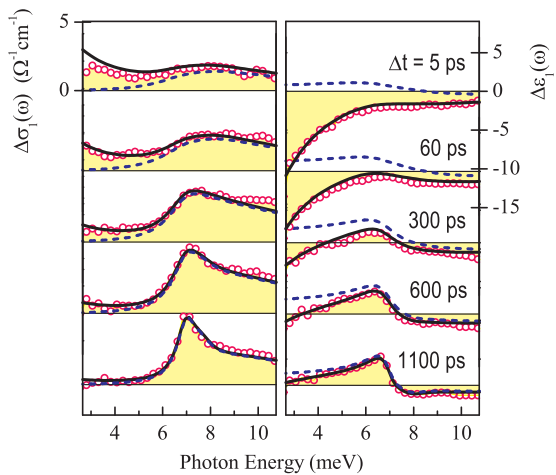}
\caption{(color online). Transient THz spectra (dots) at $T_{\rm L} = 6$~K after non-resonant excitation into the continuum (at 1.561~eV with fluence $0.2~\mu$J/cm$^{2}$). Solid lines: two-component model; dashed lines: intra-excitonic model only, with broadening $\Gamma_{bb} = 3$~meV (5~ps), 2.8~meV (60~ps), 2~meV (300~ps), 1.5~meV (600~ps), and 1~meV (1100~ps) and densities as in Figs.~\ref{Fig_FormationAnalysis}(a) and \ref{Fig_FormationAnalysis}(b).
\label{Fig_Formation_Spec}}
\end{figure}

Having verified the applicability and correct gauge of the above dielectric function for the study of mixed $e$-$h$ gases, we can now analyze the kinetics of exciton formation from unbound pairs. For this, the QWs are excited non-resonantly above the band gap. Transient THz spectra are shown in Fig.~\ref{Fig_Formation_Spec}. They exhibit a complex dynamics which evolves from a broad Drude-like response into an intra-excitonic line shape. Directly after excitation, the low-frequency conductivity in $\Delta\sigma_1$ and the $-1/\omega^2$ dispersion in $\Delta \epsilon_1$ reveal a conducting, Drude-like phase. However, a broad excitonic peak is also evident in $\Delta \sigma_1$ at these early delay times, rendering the spectra similar to the mixed ionized phase discussed above. With increasing delay the Drude component decays, while spectral weight builds up around the 1$s$-2$p$ transition until the intra-excitonic line shape is restored. Exciton formation thus proceeds on two different time scales: fast formation of a large exciton fraction directly after excitation, and a slow binding of remaining free carriers into excitons within several 100~ps.

\subsection{Formation kinetics}

\begin{figure}          
\includegraphics[width=6.3cm]{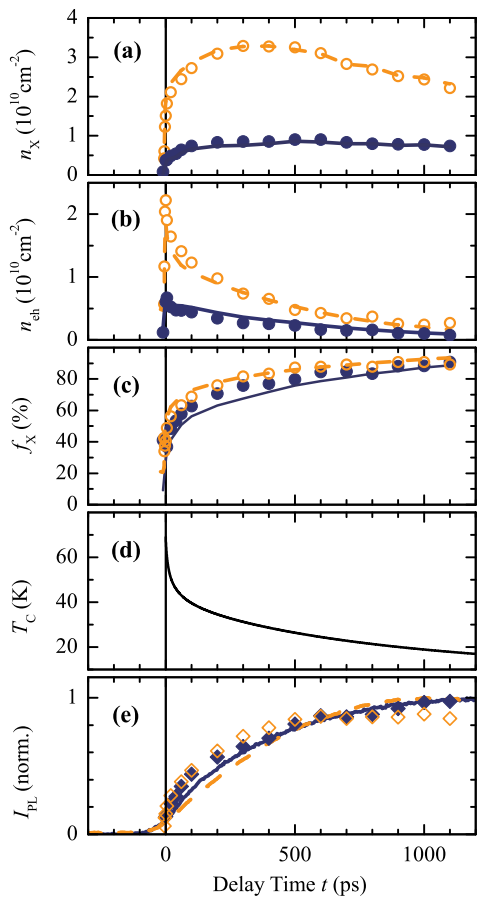}
\caption{(color online). Pair density and luminescence during exciton formation. [(a)-(c)] Exciton density, free-carrier density, and exciton fraction from the two-component model in Fig.~\ref{Fig_Formation_Spec} (solid dots) and from data at $\simeq3.5\times$ higher density (open circles). The Saha model (see text) is shown as solid and dashed lines for low and high densities, respectively. (d) Simulated carrier temperature $T_{\rm C}$ for $\alpha=3$, $T_{\rm L} = 6$~K. (e) PL intensity $I_{\rm PL}$ after non-resonant excitation at 1.561 eV ($T_{\rm L}$\,=\,4.2\,K) with densities of $1\times10^{10}$~cm$^{-2}$ (solid line) and $3.2\times10^{10}$~cm$^{-2}$ (dashed line). The PL intensity calculated from $n_{\rm X}(t)$ and $T_{\rm C}$ in panels (a) and (d) is indicated for low and high densities, respectively, as solid and open diamonds.}\label{Fig_FormationAnalysis}
\end{figure}

For quantitative insight, we analyzed the transient spectra with the two-component dielectric function (solid lines in Fig.~\ref{Fig_Formation_Spec}). The corresponding exciton and unbound pair densities and the exciton fraction are shown in Figs.~\ref{Fig_FormationAnalysis}(a)--\ref{Fig_FormationAnalysis}(c), as obtained from the above data (solid dots) and from a data set with $\simeq 3.5$ times higher density (open circles).
In both cases, the analysis confirms a fast initial generation of an appreciable exciton fraction $f_{\rm X} \approx 40 \%$, followed by slower transfer of unbound $e$-$h$ pairs into excitons and an eventual decay of the exciton density due to recombination. At the longest delays ($\Delta t \approx 1$~ns) almost all carriers are bound into excitons with $f_{\rm X} \approx 90 \%$.

After non-resonant excitation, photoexcited $e$-$h$ gases are known to thermalize on a 100-fs time scale into a Fermi distribution, which cools to the lattice temperature via emission of optical and acoustic phonons.\cite{Sha99} This raises the question whether Figs.~\ref{Fig_FormationAnalysis}(a)--\ref{Fig_FormationAnalysis}(c) reflect a
pair kinetics where ($i$) exciton formation is much faster than cooling, such that the carrier gas maintains thermodynamic equilibrium at each time, or where ($ii$) the time evolution is limited by slower, bi-molecular $e$-$h$ pairing interactions resulting in a measurable, non-equilibrium deviation from the Saha condition. To resolve this matter, we have calculated the cooling dynamics of the quasi-2D carrier temperature $T_{\rm C}(t)$ by integrating the time-dependent change
\begin{equation}
\frac{dT_{\rm C}}{dt} = - \frac{\langle dE/dt \rangle_{T_{\rm C}}}{2k_{\rm B}\alpha} + \frac{\langle dE/dt \rangle_{T_{\rm L}}}{2k_{\rm B} \alpha} \; .
\end{equation}
\noindent where the second term ensures equilibration at the lattice temperature. Here, $\langle dE/dt \rangle_T$  is the energy-loss rate of the $e$-$h$ gas at temperature $T$, for which we directly employed the values obtained for GaAs QWs by Leo {\it et al}.\cite{Leo88} Moreover, $\alpha$ is the well-known reduction factor due to hot-phonon effects, and $\alpha = 3$ was chosen in agreement with previous work.\cite{Syc04,Leo87,Leo88} Given the excitation conditions, we estimate an initial carrier temperature $T_{\rm C}(0) = 69$~K after distributing the excess energy of the photoexcited carriers equally among electrons and holes.\cite{Leo87} The resulting cooling curve is shown in Fig.~\ref{Fig_FormationAnalysis}(d). It is characterized by a quick drop to a temperature of $\simeq 40$~K due to emission of LO phonons, after which the cooling proceeds via acoustic phonons on a much longer time scale. The values and overall shape of this cooling dynamics closely agrees with previous work.\cite{Yoo96,Syc04}

We can now compare the pair densities and exciton fraction obtained from the transient THz spectra with the predictions of the Saha equation. This model description assumes that the excitons and free carriers form a quasi-equilibrium that conforms to the time-dependent carrier temperature $T_{\rm C}(t)$, via equilibration processes that are fast compared to the overall dynamics. Accordingly, the Saha equilibrium exciton fraction and densities were calculated from Eq.~\ref{Eq_Saha}, with $N = n_{\rm X}(t) + n_{\rm eh}(t)$ as obtained from the measured THz spectra and with $T = T_{\rm C}(t)$. The resulting curves are shown in Figs.~\ref{Fig_FormationAnalysis}(a)--\ref{Fig_FormationAnalysis}(c) as solid and dashed lines, clearly yielding an extremely close quantitative description of the experimentally-derived densities and exciton fraction. Hence, the thermodynamic equilibrium model provides for a surprisingly simple description of the formation kinetics.

The capability of the Saha quasi-equilibrium to explain the time-dependent exciton formation kinetics agrees with observations in several PL studies.\cite{Roba95,Yoo96} However, it should be emphasized that our observations do not rule out the influence of bi-molecular interactions in the formation kinetics.\cite{Pie97,Aman94,Syc04} Indeed, as explained by Deveaud {\it et al.},\cite{Deve05} in the $10^{10}$~cm$^{-2}$ density range applicable to our current study the Saha and bi-molecular rate-equation models are expected to be largely commensurate, while they deviate at lower densities (see Fig.~7 of Ref.~\onlinecite{Deve05}). This motivates future work with even more sensitive THz probes to explore formation in the low-intensity limit, and calculations of exciton formation via carrier-carrier interactions in microscopic models.\cite{Sian01} Moreover, other nanoscale materials can be studied via THz probes to explore size-dependent electron-phonon interaction strengths and formation rates.

\subsection{Photoluminescence dynamics}

Next, we compare the pair kinetics with time-resolved PL measured in our sample, shown as lines in Fig.~\ref{Fig_FormationAnalysis}(e) for comparable, non-resonant excitation conditions. In stark contrast to the resonantly-excited case [Fig. \ref{Fig_Recombination}(b)], the PL displays a slow rise which reaches its maximum only $\approx 1000$~ps after excitation. Such a delayed rise agrees with previous PL studies of exciton formation.\cite{Dam90,Blo93,Kum96,Yoo96,Syc04,Mari97,Bajo06} The fast initial exciton formation is largely absent in the PL, which indicates that the excitons probed by the THz pulses at early delays primarily populate high energy states with momenta $K \gg 0$.

There has been a long debate on whether exciton luminescence can arise in a plasma of unbound $e$-$h$ pairs or whether it is fully explained by ``bright'' excitons around $K=0$.\cite{Kir98,Chat04,Szc05} In this respect, our observations contrast sharply with a recent THz study that concluded the absence of THz absorption (and thus of excitons) with simultaneous observation of exciton PL after non-resonant excitation.\cite{Gal05} In that study, the THz absorption was probed around a single wavelength assumed to coincide with the $1s$-$2p$ transition. Moreover, the near-IR exciton line was almost $10\times$ broader than in our sample, pointing to significant inhomogeneities that can result in carrier localization. In contrast, the present experiments provide full and broadband spectral information of the THz response, in both real and imaginary parts. Our results, obtained in a high-quality QW sample, clearly resolve the existence of a significant THz absorption peak at the $1s$-$2p$ transition -- revealing a large exciton density -- directly after non-resonant excitation, while the PL rises only slowly.

Indeed, given the THz-derived densities we can test whether our measured PL dynamics is fully explained if we assume that only excitons contribute to the luminescence. Following previous work (see, e.g., Ref.~\onlinecite{Yoo96}), the intensity of the luminescent fraction can in this case be written as
\begin{equation}\label{Eq_PL_Intens}
I_{\rm PL}(t) \propto \frac{n_{\rm X}(t)}{\hbar \Gamma_{\rm h}} \left( 1 - e^{-\hbar \Gamma_{\rm h}/k_{\rm B} T_{\rm C}(t)}\right) \approx \frac{n_{\rm X}(t)}{k_{\rm B} T_{\rm C}(t)}
\end{equation}
\noindent where $\Gamma_{\rm h}$ is the near-IR homogeneous linewidth, and the intuitively simple approximation on the right hand side is valid for $\hbar\Gamma_{\rm h} \ll k_{\rm B} T_{\rm C}$. We verified that the shape of the PL dynamics is not affected by the approximation, by comparing it with the full expression in Eq.~\ref{Eq_PL_Intens} with a density-dependent $\Gamma_{\rm h}$.\cite{Hono89} For the calculation, we use the experimentally-derived $n_{\rm X}(t)$ [Fig.~\ref{Fig_FormationAnalysis}(a)], and the above-discussed $T_{\rm C}(t)$ [Fig.~\ref{Fig_FormationAnalysis}(d)] which resulted in a close description of $f_{\rm X}(t)$. The resulting calculated PL intensity is shown as diamonds in Fig.~\ref{Fig_FormationAnalysis}(e). It provides a good representation of the luminescence rise time, underscoring the consistent agreement between the PL and THz signals in the above scenario. Remaining differences in the calculated and measured PL shapes can be explained in part by the lower time resolution of the PL experiment, and by inherent limitations of our simplified analysis that assumes fully thermalized exciton and free-carrier distributions. This comparison quantitatively confirms the sensitivity of luminescence to optically-active excitons around $K \approx 0$, leading to the predominance of relaxation to low-energy states rather than exciton formation in the PL kinetics.

Thus, our two-component THz analysis provides a fully-consistent description of exciton and free-carrier dynamics across a large set of experimental data. The applicability of this model is underscored by: ($i$) the absolute density scaling, corroborated by excellent agreement between the absorbed photon flux and the densities obtained from model fits; with excitons and free carriers gauged separately, ($ii$) the close agreement in shape with measured spectra in both real and imaginary parts, ($iii$) the precise fulfillment of the intraband sum rule, ($iv$) the quantitative agreement of the THz-derived exciton fraction after ionization with thermodynamic equilibrium (Saha equation) at a consistent absolute pair density, ($v$) the likewise quantitative agreement of the THz-derived ionization rate with the known LO-phonon ionization frequency, ($vi$) the pair kinetics during exciton formation which follows closely a Saha equilibrium, a physically reasonable result assuming rapid formation at these densities, and finally as explained above ($vii$) the consistent and quantitative description of PL dynamics using the THz-derived pair densities.

\section{Conclusions}\label{Sec_Conclusion}

To conclude, we discussed optical-pump THz-probe studies and a detailed model analysis of the transient THz spectra and density kinetics of quasi-2D excitons and unbound $e$-$h$ pairs in GaAs quantum wells. An intra-excitonic dielectric function was presented, whose shape and absolute density scaling is in excellent agreement with the measured THz response of resonantly-generated excitons. Ionization of excitons, in turn, leads to THz spectra that exhibit both intra-excitonic and Drude-like features. Here, a two-component dielectric function successfully describes the complex THz spectra and yields densities of excitons and unbound $e$-$h$ pairs. Ionization is found to result in a quasi-equilibrium, whose exciton fraction quantitatively agrees with the Saha thermodynamic equilibrium -- thus experimentally verifying the density scaling of the two-component model.

The analysis is equally well applied to transient THz spectra during exciton formation, demonstrating fast initial formation of~$\simeq40\%$ excitons followed by slower pair binding within several 100~ps. At the longest delays, about~$90\%$ of the pairs are bound into excitons. The time-dependent exciton fraction at our densities is quantitatively described by a Saha equilibrium that follows the $e$-$h$ gas cooling dynamics. Finally, in this scenario a consistent agreement is found with time-resolved PL measured for comparison. Our study provides the basis for further exploration of the pair kinetics of quasi-2D $e$-$h$ gases via transient THz spectroscopy and holds promise for enabling new studies of exciton physics inaccessible with near-IR light.

\begin{acknowledgments}
\noindent We thank R.~L\"{o}venich for contributions in the early stages of this work and J. Reno for providing quantum well samples. Our investigations were supported by the Director, Office of Science, Office of Basic Energy Sciences, of the U.S. Department of Energy under Contract No. DE-AC02-05CH11231.
\end{acknowledgments}

\vspace{5 mm}

\appendix

\section{Complex THz Transmission Function of Multiple Quantum Wells}\label{Sec_EvalAppendix}

\begin{figure}           
\includegraphics[width=8.4cm]{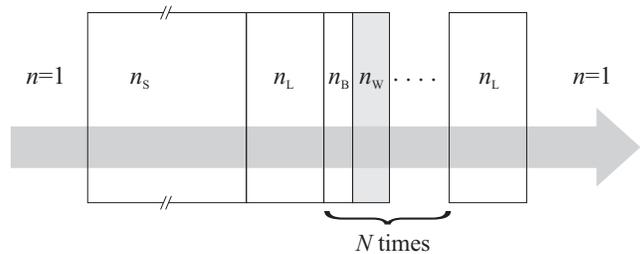}
\caption{Multi-layer geometry consisting of a thick substrate (S), spacer layer (L), a thin stack of $N$ alternating barrier (B) and well (W) layers, and second spacer layer.\label{Fig_Eval_1}}
\end{figure}

\noindent In the following, we deduce the complex THz transmission function (and its pump-induced change) for a multiple-quantum-well structure. Figure~\ref{Fig_Eval_1} shows the geometry, modeled as a thin multi-layer stack on a thick dielectric substrate. This treatment, rather than a single-layer approximation, becomes relevant at the lowest THz frequencies, since each absorbing layer adds a Fresnel phase shift that scales approximately inversely with frequency. We can write the complex transmission of the layered system using a matrix approach\cite{Bor99}
\begin{equation}\label{Eq_multil_full}
t(\omega) = \frac{4 n_{\rm S}}{(S_{11}+S_{12})n_{\rm S} + S_{22} + S_{21}}\cdot \frac{e^{i \frac{\omega}{c} n_{\rm S} d_{\rm S}}}{n_{\rm S} + 1} \end{equation}

\noindent where $\mathbf{S}$ is a matrix describing the response of the multi-layer stack, while $n_{\rm S}$ and $d_{\rm S}$ are the substrate refractive index and thickness, respectively. This expression takes into account all reflections within the thin stack but ignores multiple reflections in the substrate, since our electro-optically sampled THz field trace covers only the first pulse replica. For the above geometry $\mathbf{S} \equiv \mathbf{L} \cdot \mathbf{M} \cdot \mathbf{L}$, where the matrix
\begin{equation}
\mathbf{L}  = \left(\begin{array}{ccc}
     \cos \beta_{\rm L} & -\frac{i}{n_{\rm L}} \sin \beta_{\rm L}  \\
     -i n_{\rm L} \sin \beta_{\rm L} & \cos \beta_{\rm L}\end{array} \right)
\end{equation}

\noindent corresponds to each of the two spacer layers and
\begin{equation}
\mathbf{M}  = \left(\begin{array}{ccc} M_{11} & M_{12}\\ M_{21} & M_{22}\end{array} \right)
\end{equation}

\noindent describes the multi-layer sequence of quantum wells and barriers, with
\begin{eqnarray}
\nonumber M_{11} & = & \mathcal{U}_{N-1}(a) \left(\cos \beta_{\rm B} \cos \beta_{\rm W} - \frac{n_{\rm W}}{n_{\rm B}}\sin \beta_{\rm B} \sin \beta_{\rm W} \right)\\
\nonumber & & - \mathcal{U}_{N-2}(a)\\
\nonumber M_{12} & = & -i \, \mathcal{U}_{N-1}(a)\left(\frac{\cos \beta_{\rm B} \sin \beta_{\rm W}}{n_{\rm W}} + \frac{\sin \beta_{\rm B} \cos \beta_{\rm W}}{n_{\rm B}}\right)\\
\nonumber M_{21} & = & -i \, \mathcal{U}_{N-1}(a) \times \\
\nonumber & & \;\;\;\;\;\;\;\; (n_{\rm B} \sin \beta_{\rm B} \cos \beta_{\rm W} + n_{\rm W} \cos \beta_{\rm B} \sin \beta_{\rm W})\\
\nonumber M_{22} & = & \mathcal{U}_{N-1}(a) \left( \cos \beta_{\rm B} \cos \beta_{\rm W} - \frac{n_{\rm B}}{n_{\rm W}}\sin \beta_{\rm B} \sin \beta_{\rm W} \right)\\
& & -\mathcal{U}_{N-2}(a)
\end{eqnarray}

\noindent In the above, $N$ is the number of  barrier-QW layer pairs, while $n_i$ is the refractive index and $d_i$ is the thickness of layer $i$ (with $i$ = W, B, L denoting well, barrier, and spacer layer, respectively). Moreover, $\beta_i \equiv \frac{\omega}{c}n_i d_i$ are phase factors, and $\mathcal{U}_{N}(x)= (1-x^2)^{-1/2}\sin[(N+1)\arccos(x)]$ are the Chebyshev polynomials of the second kind, \cite{Bor99} with

\begin{equation}
a \equiv \cos \beta_{\rm B} \cos \beta_{\rm W} - \frac{1}{2}(\frac{n_{\rm B}}{n_{\rm W}}+\frac{n_{\rm W}}{n_{\rm B}})\sin \beta_{\rm B} \sin \beta_{\rm W}.
\end{equation}

\noindent Approximating the formulas for our case of optically-thin media, $\beta_i \ll 1$, yields $a \approx 1$ and $\mathcal{U}_{N}(a) \longrightarrow N +1$. Terms of the order of $\beta_i^2$ and higher can be accordingly neglected, such that
\begin{equation}\label{Eq_N_simple}
\mathbf{L}  \approx \left(\begin{array}{ccc}
     1 & -i\frac{\omega}{c} d_{\rm L}\\
     -i \frac{\omega}{c} d_{\rm L} n_{\rm L}^2 & 1\end{array} \right)
\end{equation}
\begin{equation}\label{Eq_M_simple}
\mathbf{M}  \approx \left(\begin{array}{ccc} 1
& -iN \frac{\omega}{c}(d_{\rm B} + d_{\rm W})\\
-iN \frac{\omega}{c}(d_{\rm B} n_{\rm B}^2 + d_{\rm W} n_{\rm W}^2) & 1
\end{array} \right)
\end{equation}

\noindent Combining Eqs.~\ref{Eq_multil_full},~\ref{Eq_N_simple}, and \ref{Eq_M_simple}, and neglecting terms of the order of $(\frac{\omega}{c} d_{\rm L} n_{\rm L})^2$ and $(\frac{\omega}{c} d_{\rm B} n_{\rm B})^2$ or higher yields the complex THz transmission coefficient
\begin{equation}
t(\omega) = \frac{4 n_{\rm S}}{1+ n_{\rm S} - \epsilon(\omega) A + B} \cdot \frac{e^{i \beta_{\rm S}}}{n_{\rm S} + 1}\label{Eq_multil_transmission}
\end{equation}

\noindent with
\begin{eqnarray}
A & \equiv & i\frac{\omega}{c}D_{\rm QW}[1-i\frac{\omega}{c}d_{\rm L}(1+n_{\rm S})]\nonumber\\
B & \equiv & -i\frac{\omega}{c}(n_{\rm S} D_{\rm tot} + 2d_{\rm L} n_{\rm L}^2 + N d_{\rm B} n_{\rm B}^2)\nonumber
\end{eqnarray}

\noindent where $D_{\rm tot} \equiv 2d_{\rm L} + N(d_{\rm B} + d_{\rm W})$ is the total multi-layer thickness and $D_{\rm QW} \equiv N d_{\rm W}$ is the aggregate thickness of quantum well material. In the above, $\epsilon(\omega) \equiv n_{\rm W}^2(\omega)$ is the complex-valued quantum-well dielectric function. Note that in the limit $d_{\rm L}$,~$d_{\rm B} \rightarrow 0$, the above expression reduces to the transmission of a single optically-thin, absorbing layer on a dielectric substrate.

In equilibrium, Eq.~\ref{Eq_multil_transmission} directly connects the static dielectric function $\epsilon(\omega)$ to the experimentally-accessible complex transmission coefficient given by $t(\omega)=E(\omega)/E_{\rm in}(\omega)$. Here, $E_{\rm in}(\omega)$ and $E(\omega)$ are the incoming and transmitted fields, respectively. Likewise, in the photoexcited state the modified dielectric function $\epsilon(\omega) + \Delta\epsilon(\omega)$ is linked to $t^*(\omega) = [E(\omega)+\Delta E(\omega)]/E_{\rm in}(\omega)$, where $\Delta E(\omega)$ is the pump-induced field change. The ratio of these transmission coefficients is then given by
\begin{equation}
\frac{t(\omega)}{t^*(\omega)} = \frac{E(\omega)}{E(\omega) + \Delta E(\omega)}. \label{Eq_tratio}
\end{equation}

\noindent Combined with Eq.~\ref{Eq_multil_transmission}, this yields an analytical expression for the pump-induced \textit{change} in the quantum well THz dielectric function,
\begin{equation}
\Delta\epsilon(\omega) = \frac{\Delta E(\omega)}{E(\omega) + \Delta E(\omega)} \cdot \left(\frac{1+n_{\rm S}+B}{A} -\epsilon(\omega)\right).
\label{Eq_Dn_multi}
\end{equation}
\vspace{1mm}

\noindent The above multi-layer expression accounts for the effects of Fresnel phase shifts at each interface. For our specific structure, we have $N = 10$, $d_{\rm L} = 500$~nm, $d_{\rm B} = 10$~nm, $d_{\rm W} = 14$~nm, $n_{\rm S} =$3.1, $n_{\rm L} = n_{\rm B} =$3.4, and (in equilibrium) $n_{\rm W} =$3.6. While Eqs.~\ref{Eq_multil_transmission} and \ref{Eq_Dn_multi} are sufficient approximations for most conditions, the dielectric function change $\Delta\epsilon(\omega)$ in the high-density regime can be obtained via numerical solution of Eqs.~\ref{Eq_multil_full} and \ref{Eq_tratio}.

\section{Intra-Exciton Terahertz Dielectric Response}\label{Sec_ExcitonTheoryAppendix}

\noindent Below, we provide a detailed derivation of the THz dielectric response in Eq.~\ref{Eq_IntraExcitonDielecFunc} due to transitions from the $1s$ exciton state into higher bound states and into the continuum. Particular attention is paid to scaling the response in absolute units. The bound 2D exciton normalized wave functions are\cite{Shi66,Eke87,Hau04}
\begin{eqnarray}
\nonumber \psi_{n,m}(r,\phi) = \sqrt{\frac{2}{a^2 (n+\frac{1}{2})^3}\cdot\frac{(n-|m|)!}{((n+|m|)!)^3}}\\
\times \rho^{|m|} e^{-\rho/2} ~~ L_{n+|m|}^{2|m|}(\rho) ~~ \frac{e^{i m \phi}}{\sqrt{2\pi}}~,
\end{eqnarray}

\noindent where $n = 0,1,2\ldots$ indicates the main quantum number, $m$ is an integer with $|m| < n$, $L_{q}^{p}(\rho) \equiv \sum_{\nu =0}^{q-p} (-1)^{\nu+p}\rho^{\nu} (q!)^2/[\nu!(q-p-\nu)!(p+\nu)!]$ are the associated Laguerre polynomials, and $\rho \equiv 2r/((n+1/2)a)$. The Bohr radius $a$ and binding energies $E_n$ scale as in Eqs.~\ref{Eq_bohr_radius} and \ref{Eq_binding_energy}. Following Ref.~\onlinecite{Eke87}, the finite well size is taken into account by rescaling the Coulomb potential with a parameter $\lambda$, in order to recover a realistic binding energy. The quantum number $n$ is enumerated starting from zero, while we will colloquially call the ground state "$1s$" (corresponding to $n=0, m=0$), and the higher bound states accordingly "$n+1$" levels, i.e., the "$2p$" level corresponds to $n=1,m=\pm 1$.

The wave functions of unbound $e$-$h$ pairs in the continuum are, in turn,

\begin{eqnarray}
\nonumber \psi_{k,m}(r,\phi) = \frac{(2kr)^{|m|}}{(2|m|)!} F\left(|m|+\frac{1}{2}+\frac{i}{ak};2|m|+1;2ikr\right) \\
\nonumber \times e^{-ikr}~\frac{e^{i m \phi}}{\sqrt{2\pi}}
\sqrt{\frac{2k}{1+e^{-\frac{2\pi}{ak}}} \prod_{j=1}^{|m|}\left[(j-\frac{1}{2})^2+\frac{1}{a^2k^2} \right]}~, \\
\end{eqnarray}

\noindent where F(a;b;z) denotes the confluent hypergeometric function, and the product is replaced by unity for $m=0$. These functions are normalized per unit momentum, $\langle\psi^*_{k',m'}|\psi_{k,m}\rangle = \delta(k-k')\delta(m-m')$. The scalar wave number $k$ is defined via $\hbar k \equiv \sqrt{2 \mu E(k)}$, where $E(k)$ is the unbound $e$-$h$~pair kinetic energy relative to the band edge.

Using Fermi's golden rule, the contribution to the dielectric function $\epsilon(\omega)$ for excitons in the $1s$ ground state $\psi_{0,0}$ is given by
\begin{eqnarray}
\nonumber \Delta \epsilon(\omega) = \frac{n_{\rm X} e^2}{d_{\rm W} \epsilon_0 \mu} \bigg( \sum_n \frac{f_{1s,n}}{([\frac{E_n-E_0}{\hbar}]^2-\omega^2)-i\omega\Gamma_{\rm bb}} \\
+ \int \frac{f_{1s}(k)}{([\frac{E_0 + E(k)}{\hbar}]^2 - \omega^2)-i\omega\Gamma_{\rm bc}} \,\, dk \bigg) \;,
\end{eqnarray}

\noindent which represents the sum over all final states $\psi_{\rm f}$ with oscillator strengths
\begin{equation}
f_{1s,{\rm f}} \equiv \frac{2\mu}{\hbar^2}(E_{\rm f} - E_0) \,\, |\langle \psi_{\rm f} | \, \hat{x} \, | \psi_{0,0}\rangle|^2.
\end{equation}

\noindent In the above, $n_{\rm X}$ is the $1s$ exciton sheet density, $\mu$ is the reduced $e$-$h$ pair effective mass, and $\Gamma_{\rm bb}$, $\Gamma_{\rm bc}$ are phenomenological broadening parameters for bound-bound and bound-continuum transitions, respectively.

Consider first transitions to $p$-like bound exciton states. For an in-plane THz field linearly polarized along $\hat{x} = r \cos(\phi)$, the dipole matrix elements for each of two-fold degenerate final states ($m=\pm1$) are
\begin{eqnarray}
\langle\psi_{n,\pm1}| \hat{x} |\psi_{0,0}\rangle ~= ~~~~~~~~~~~~~~~~~~~~~~~~~~~~~~~~~~~~~~~~~~~~~~~~~ \\
\nonumber ~~~~\frac{a}{4} \sqrt{2~\frac{(n+\frac{1}{2})^3(n-1)!}{((n+1)!)^3}}
\times \int_0^\infty  \rho^3 e^{-(n+1)\rho} L_{n+1}^2(\rho) ~d\rho
\end{eqnarray}

\noindent The latter integral can be written analytically as
$\int_0^\infty  \rho^3 e^{-(n+1)\rho} L_{n+1}^2(\rho) ~d\rho = {(n-1)! n^n (2n+1)(n+1)^{-n-1}}$, yielding the oscillator strengths
\begin{equation}
f_{1s,n} = \frac{2 \mu a^2}{\hbar^2} (E_n - E_0) \frac{(n+\frac{1}{2})^5 n^{2n-3}}{(n+1)^{2n+5}}
\end{equation}
\noindent when including both $m=\pm1$ final states. \noindent The matrix elements for transitions from $1s$ into $p$-like continuum states are, in turn, given by
\begin{eqnarray}
\nonumber <\psi_{k,\pm1}| \hat{x} |\psi_{0,0}> = \frac{2k}{a} \sqrt{\frac{2k}{1+e^{-\frac{2\pi}{ak}}} (\frac{1}{4}+\frac{1}{a^2k^2})}~~~~~~~~\\
\nonumber \times \int_0^\infty e^{-\frac{2r}{a}-ikr} r^3 ~F\left(\frac{3}{2}+\frac{i}{ak};3;2ikr\right) dr\\
\end{eqnarray}
\noindent The above integral can be written as $\int_0^\infty e^{-2r/a-ikr} r^3 ~F({3\over2}+\frac{i}{ak};3;2ikr) dr
= ~-8 i a^4 (1+2ak/(2i-ak))^{-\frac{5}{2}-\frac{i}{ak}}(-2i+ak)^{-5}$, yielding the oscillator strengths per unit momentum
\begin{eqnarray}
\nonumber f_{1s}(k) = \frac{2 \mu a^2}{\hbar^2} \left(E_0+E(k)\right) \frac{256 a^2 k\, (\frac{2i+ak}{2i-ak})^{-\frac{2i}{ak}}}{(1+e^{-\frac{2\pi}{ak}})(4+a^2k^2)^4}\\
\end{eqnarray}

\noindent as the sum of transitions to $m=\pm1$. With the above, we obtain the dielectric response $\Delta\epsilon(\omega)$ in Eq.~\ref{Eq_IntraExcitonDielecFunc} arising from intra-excitonic transitions between the $1s$ state and all higher bound states and the continuum.

Finally,  we can verify the expected smooth transition of the dielectric function at the edge between bound and continuum transitions, in analogy to the intra-atomic absorption of hydrogen.\cite{Som47} The theory used to derive the 2D exciton wave functions links the quantum number $\tilde{n}$ and wave number $k$ via $\tilde{n} + {1\over2} = 1 / (ia_0k)$. For bound states, $\tilde{n}$ is integer and real, while for continuum states it is a continuous and imaginary number. To compare the two components in Eq.~\ref{Eq_IntraExcitonDielecFunc}, the oscillator strengths must be normalized to a constant "quantum number" interval $\Delta \tilde{n} =1$, i.e., the $k$-dependent dipole moment must be multiplied by $(dk/dn)\Delta n = -1/(ia_0 (\tilde{n}+{1\over2})^2)$. Substituting $ 1 / (ia_0k) \longrightarrow \tilde{n} + {1\over2}$ then transforms the bound-continuum expression into one that differs from the bound-bound expression only by a factor of $(1+e^{-2\pi|n+{1\over2}|})^{-1}$. Since the latter approaches unity for $n \rightarrow \infty$, both expressions analytically match up smoothly at the continuum edge. This fact is illustrated in Fig.~\ref{Fig_THzDrudeExciton}(g) by numerical simulation.

\bibliography{D:/Publications/Bibtex_Library/Kaindl_BibTex_Refs}
\end{document}